# Up-conversion detection of mid-infrared light carrying orbital angular momentum

Zheng Ge[1, 2, 3], Chen Yang[1, 2, 3], Yin-hai Li[1, 2, 3], Yan Li[1, 2, 3], Shi-Kai Liu[1, 2, 3], Su-Jian Niu[1, 2, 3], Zhi-Yuan Zhou[1, 2, 3, *], and Bao-Sen Shi[1, 2, 3, *]

[1] *CAS Key Laboratory of Quantum Information, University of Science and Technology of China, Hefei 230026, China*

[2] *CAS Center for Excellence in Quantum Information and Quantum Physics, University of Science and Technology of China, Hefei 230026, China*

[3] *Hefei National Laboratory, University of Science and Technology of China, Hefei 230088, China*

*\* zyzhouphy@ustc.edu.cn*

*\* drshi@ustc.edu.cn*

Frequency up-conversion is an effective method of mid-infrared (MIR) detection by converting the long-wavelength photons to the visible domain, where efficient detectors are readily available. Here, we generate the MIR light carrying orbital angular momentum (OAM) from a difference frequency generation process and perform the up-conversion of it via sum frequency conversion in a bulk quasi-phase-matching crystal. The maximum quantum conversion efficiencies from MIR to visible are 34.0%, 10.4%, and 3.5% for light with topological charges of 0, 1, and 2, respectively, which is achieved by utilizing an optimized strong pump light. We also verify the OAM conservation with a specially designed interferometer, and the results agree well with the numerical simulations. Our study opens up the possibilities for generating, manipulating, and detecting MIR light that carries OAM, and will have great potential for optical communications and remote sensing in the MIR regime.

## 1. Introduction

The mid-infrared (MIR) band covers the absorption spectrum of many molecules [1] and is closely related to the thermal radiation of objects, which has been used in many aspects, such as environmental monitoring [2–4], geology for mineral identification [5], stand-off detection [6], and biomedical science [7–10]. Among them, the 3-5 micron band corresponds to one of the atmospheric communication windows, which is of potential importance in remote sensing [11] and communication fields [12,13]. On the other hand, light that carries orbital angular momentum (OAM) has stimulated considerable research interest in both classical and quantum optical domains [14–21]. This particular beam with an azimuthal phase $\exp(il\phi)$ is well-known as owning an exact OAM of $l\hbar$ per photon [22], where $l$ and $\phi$ refer to the topological charge (TC) and azimuthal angle respectively. MIR light that carries OAM is of great value in many specific applications, such as enhancing the information channel capacity in communication [23–25] and helping the understanding and formation of chiral microstructures [26,27]. Whereas the detection equipment in the MIR band is not mature at present, reflecting on the lower detection sensitivity, higher noise, and narrower bandwidth compared with its visible or near-infrared (NIR) counterpart. Consequently, it is more effective to detect the MIR light after converting it into visible/NIR light, utilizing high-performance detectors based on wide bandgap materials like silicon [28]. Due to the high effective nonlinear coefficient and elimination of the walk-off effect, the QPM technique has been used extensively for frequency conversion of light carrying OAM in previous works [29–31]. Up to now, the effective up-conversion of mid-infrared light has been realized by using a waveguide [32,33]. However, compared with the traditional bulk crystals, the loss of spatial information makes the waveguide-based up-conversion unable to meet a wider range of detection requirements. In addition, the current waveguide-based nonlinear transformation is mainly in the single-mode case, while higher-order mode frequency conversion still faces some difficulties. Bulk crystals have been widely used in many practical applications of frequency conversion because they can keep the phase and spatial information during the nonlinear process [34–40]. In this case, however, the beam waist radius in the center of the crystal is larger than that in a waveguide, which asks for a much higher pump power to improve the quantum conversion efficiency (QCE). In previous works on MIR up-conversion detection, cavity-enhanced [41,42] or pulsed light pumped [43–45] systems were employed, achieving satisfactory QCE. However, a systematic study of the frequency conversion of OAM modes in the MIR band has not been reported to date.

In this work, the cascaded frequency conversion of light carrying OAM from 792 nm to 3100 nm and back to 792 nm was demonstrated, pumped by a high power continuous-wave (CW) light. The laser at 3100 nm was generated from a difference frequency generation (DFG) process, serving as the MIR source. Two identical MgO-doped periodically poled lithium niobate (MgO: PPLN from Covesion Ltd.) bulk crystals were utilized in the nonlinear processes above, each of which has a length of 40 mm and has nine poling periods ranging from 20.9 to 23.3 μm in steps of 0.3 μm. With the temperature of the crystals controlled, the nonlinear processes satisfied the type-0 quasi-phase-matching condition. Here we used the channel with a poled period of 20.6 μm and an aperture of 0.5 mm by 0.5 mm. For the convenience of discussion, in both three-wave mixing processes, the respective wavelengths were defined as $\lambda_s = 792$ nm, $\lambda_p = 1063.8$ nm, and $\lambda_i = 3100$ nm, satisfying the relation $1/\lambda_s = 1/\lambda_p + 1/\lambda_i$. Based on the nonlinear coupling equations, we proposed an analytical expression in un-depleted approximation, which described the up-conversion efficiencies for various OAM values. Meanwhile, in the case of the depleted condition, the results given by numerical calculations were presented and made a comparison with the experimental results. The final power efficiencies realized for conversion from MIR to visible are 133.1%, 40.7%, and 13.6% for TC of $l = 0$, 1, and 2, respectively, and the corresponding maximum QCEs are 34.0%, 10.4%, and 3.5%. We also showed that the OAM is conserved in the conversion process. The high conversion efficiency and well-preserved phase information indicated that our primary study for MIR up-conversion is both reliable and useful, and will pave the way for further applications in remote sensing, high capacity optical communications, and image detection.

## 2. Theoretical model

The theoretical analysis for SFG, which is based on a second-order nonlinearity, is shown as follows. The nonlinear process involves the mixing of three waves, including a strong pump wave at frequency $\omega_p$, an idler wave to be converted at frequency $\omega_i$, and the up-converted beam at frequency $\omega_s$. In our experiment, the pump light is a normal Gaussian beam, while the idler light is in the OAM mode with a TC of $l$. In the un-depleted pump approximation, the nonlinear coupled equations can be simplified as [46]:

$$\begin{cases} \dfrac{2in_i\omega_i}{c}\dfrac{dA_i}{dz} + \nabla_\perp^2 A_i = -\dfrac{d_{eff}\omega_i^2}{\varepsilon_0 c^2} A_p^* A_s \exp(-i\Delta k z) \\ \dfrac{2in_s\omega_s}{c}\dfrac{dA_s}{dz} + \nabla_\perp^2 A_s = -\dfrac{d_{eff}\omega_s^2}{\varepsilon_0 c^2} A_p A_i \exp(i\Delta k z) \end{cases} \quad (1)$$

where $d_{eff}$ is the effective nonlinear efficiency of the crystal; $\varepsilon_0$ is the permittivity of vacuum; $n_j (j = p, i, s)$ are the refractive index inside the crystal and the subscripts correspond to the pump, the idler, and the signal light respectively; $\Delta k = k_s - k_i - k_p + 2\pi/\Lambda$ is the phase mismatch in the SFG process and $\Lambda$ is the poling period of crystal; $A_j (j = p, i, s)$ are electrical fields of the pump, the idler and the signal beams, which can be expressed as [47]:

$$A_j(r,z) = \sqrt{\frac{P_j}{\pi l! \varepsilon_0 n_j c}} \frac{(\sqrt{2}r)^l}{[\omega_{0j}(1+\frac{iz}{Z_{0j}})]^{l+1}} \exp[-\frac{r^2}{\omega_{0j}^2(1+\frac{iz}{Z_{0j}})}]\exp(il\phi) \quad (2)$$

where $n_j (j=p,i,s)$ are the refractive indexes of the pump, the idler, and the signal beams inside the crystal; $\omega_{0j} (j=p,i,s)$ are the beam waists; $Z_{0j} = \pi n_j \omega_{0j}^2 / \lambda_j$ $(j=p,i,s)$ are the Rayleigh ranges of these beams; $l$ refers to the value of TC and is equal to zero in Gaussian mode; and $\phi = \tan^{-1}(\frac{y}{x})$. We have directly omitted the term containing Gouy phase shift here, which can be ignored since the two input beams have approximately matching phases according to our experimental conditions. When considering the slowly varying amplitude approximation and the un-depleted pump approximation, an analytical expression of the SFG power can be obtained as follows [48]:

$$P_s = \frac{16\pi^2 d_{eff}^2 P_p P_i L 2^l}{\varepsilon_0 c n_s n_i \lambda_s^2 \lambda_p} h(l,\xi) \quad (3)$$

where $L$ is the length of the crystal; $P_j (j=p,i,s)$ are the pump, the idler, and the signal power of these beams; $h(l,\xi)$ is the focusing function defined as:

$$h(l,\xi) = \frac{1}{\xi}\int_{-\xi}^{\xi} dx \int_{-\xi}^{\xi} dy \frac{(1+ix)^l(1-iy)^l e^{i\sigma(x-y)}}{\{(1+ix)(1-iy)[2+\frac{i(x-y)}{\beta}]+\alpha(1+\frac{ix}{\beta})(1-\frac{iy}{\beta})[2+i(x-y)]\}^{l+1}} \quad (4)$$

$\xi = L/b_p$ is defined as the focusing parameter of the pump beam, where $b_p = 2Z_{0p}$ is the confocal parameter; $\alpha = w_{0s}^2/w_{op}^2$ and $\beta = b_i/b_p$ are determined by the waist ratio of the two beams; $\sigma = \Delta k b_p/2$ is the phase-mismatching parameter. Obviously, the loss of both pump light and input idler light is ignored when obtaining the analytical expressions, which may produce deviations in specific experiments. More discussion will be presented in the subsequent analysis of the experimental results. Therefore, numerical simulation was also conducted based on the coupled wave equations, utilizing a technique called the split-step Fourier method [49,50]. The basic assumption here is that spatial evolution and nonlinear effects can act separately for each small distance traveled by the light field during transmission. In this case, the transfer process from $z$ to $z+dz$ can be carried out in two steps. In the first step, only non-linear effects are considered in Eq. (1), which gives:

$$A_s(z+dz) = i\frac{\omega_s d_{eff}}{2\varepsilon_0 n_s c} A_i A_p e^{i\Delta k z} dz + A_s(z) \quad (5)$$

In the second step, there is only space evolution, and the Fourier transform term of the light field satisfies the following relation:

$$\mathscr{F}\{A_s(z+dz)\} = \mathscr{F}\{A_s(z)\}\exp\left(i\frac{(k_x^2+k_y^2)}{2n_3 k_3}\right) \quad (6)$$

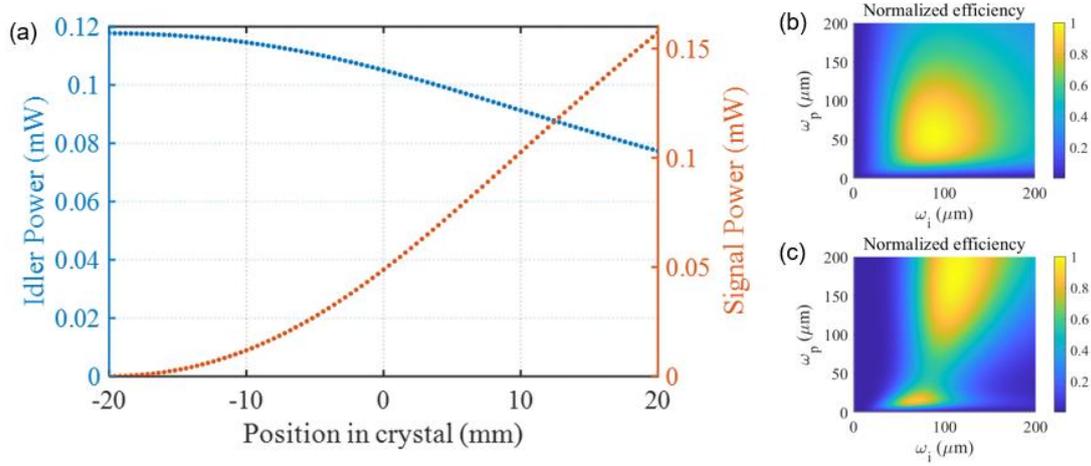

Fig. 1. (a) Dependence of the powers of the idler and signal beams on the propagation distance inside the nonlinear crystal. (b) and (c) The normalized efficiency with different beam waists for $l = 0$ and 2.

With this method, we obtained a series of discrete points after setting the initial conditions, showing the intensity variations of the idler and signal light at different positions in the crystal for Gauss mode, as shown in Fig. 1. Obviously, the accuracy of the simulation depends on the choice of step size, which also affects the speed of the calculation. An important advantage of the split-step Fourier method is that it simulates the beam mode field evolution process, which is useful for analyzing the effect of beams overlapping on the non-linear efficiency. Taking the signal light in Gauss mode and OAM mode with $l = 2$ as examples, we showed in Fig. 1(b) and (c) the predicted normalized efficiency at different beam waists, helping to find the best focusing parameters for subsequent experiments.

### 3. Experimental setup

The schematic of the experimental setup is illustrated in Fig. 2. The signal beam for the down-conversion came from a diode laser (Toptica, pro design, Graefelfing); its spatial mode was later optimized by passing through a single-mode optical fiber. The pump beam was provided by a Yb-doped fiber laser working at 1064 nm, enhanced by a fiber amplifier, and then separated into two channels, pumping the DFG and SFG modules respectively. Each laser beam was set to be vertical polarization by the wave plates before the nonlinear crystal, satisfying the restriction of phase-matching condition. A vortex phase plate (VPP) was placed before the focusing lens, imprinting OAM on the signal beam. In the first crystal, the waist sizes for the pump and the signal beams were 43 and 37 μm at the focus, respectively. The temperature of each crystal was controlled using a homemade semi-conductor Peltier temperature cooler, the temperature stability of which is ±2 mK. At the end of the DFG module, a long-pass filter removed the pump and signal beam, further before the idler beam measured by a mercury telluride detector (MCT). In the second frequency conversion process, the pump and idler beam overlapped after a dichromatic mirror, focused by the lens with beam waists of 65 μm and 110 μm respectively. The filter after the SFG crystal removed all the off-target beams but the up-converted beam at 792 nm before it entered the interference module. The input light with an OAM state of $|l\rangle$ would be converted into the form of $|l\rangle + e^{i\varphi}|-l\rangle$ by the specially designed balanced interferometer, as discussed in our previous work [51]. The result of the interference presented a petal-like pattern and was captured by a charge-coupled device camera (CCD) placed on the output side. The petals have a count of exactly $2l$, which signified that the value of the TC carried by the generated beam can be told by simply analyzing the patterns.

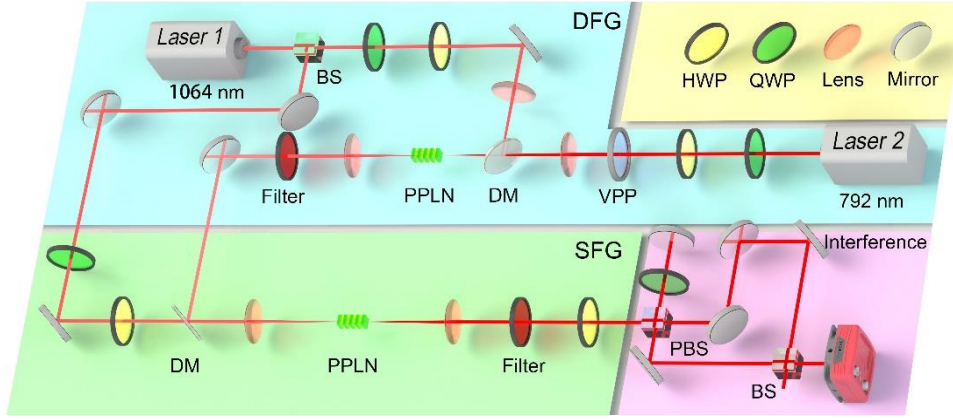

Fig. 2. Experimental setup. VPP: vortex phase plate; DM: dichromatic mirror; BPF: band-pass filter; PPLN: periodically poled lithium niobate crystal; HWP (QWP): half-wave plate (quarter-wave plate); PBS (BS): polarization beam splitter (beam splitter); CCD: charge-coupled device camera.

## 4. Results and discussion

In the first DFG module, both the input beams we used had a power of 1 W, preparing a 3100-nm Gaussian beam with a power of 2.36 mW. For ease of comparison, the idler power was adjusted to 0.2 mW with an optical attenuator for different OAM. The final power of the wave to be converted was 0.118 mW at the incident face of the crystal, suffering a total loss of 41.1% during the transmitting procedure, which was mainly brought by the dichromatic mirror because of the mismatch of the center wavelength. For varying upconversion pump power (while the idler power was maintained at 0.118 mW), the results of generated signal power for each OAM are illustrated in Fig. 3(a). Notice that the results given by the analytical calculations agree well with the experimental values initially, but gradually deviate as the pumping power increases. This deviation is not a surprise, as the small-signal approximation was used in obtaining the analytical expressions, which requires a low conversion efficiency. As the pump power increased, the consumption of MIR photons intensified and deviations between theoretical and experimental results were inevitable. The numerical calculation, on the other hand, avoided this problem and gave theoretical predictions that are relatively close to the experimental values. For both the $l = 1$ and 2 cases, there was some deviation between the theoretical and experimental values. Because the Gaussian light passing through the VPP was not in an exact Laguerre Gaussian mode [52,53], and the aberration of the MIR beams carrying OAM generated by the DFG progress was unavoidable. Considering the 2.03% power loss caused by the subsequent filter, the power efficiencies of the SFG system determined using $\eta_{power} = P_{792}/P_{3100}$ are 133.1%, 40.7%, and 13.6% for TC $l$ values of 0, 1, and 2 respectively with a pump power of 37 W, and the corresponding quantum conversion efficiencies defined by $\eta_{quantum} = \eta_{power} \lambda_{792}/\lambda_{3100}$ were 34.0%, 10.4%, and 3.5%. The conversion efficiency was satisfied in Gauss mode but reduced rapidly for increasing OAM orders. The main cause was different overlaps between the idler and the pump beams, as the OAM charge would affect the beam size and amplitude vividly. Consequently, in the up-conversion of structured beams with different OAM modes, the focusing parameters can be adjusted utilizing the same method shown in Fig. 1 (b) and (c), which would optimize the efficiency of SFG to some extent. To further eliminate the dependence of the conversion efficiency on the TC, modulation methods such as flat-top pump or imaging techniques can be considered [54,55].

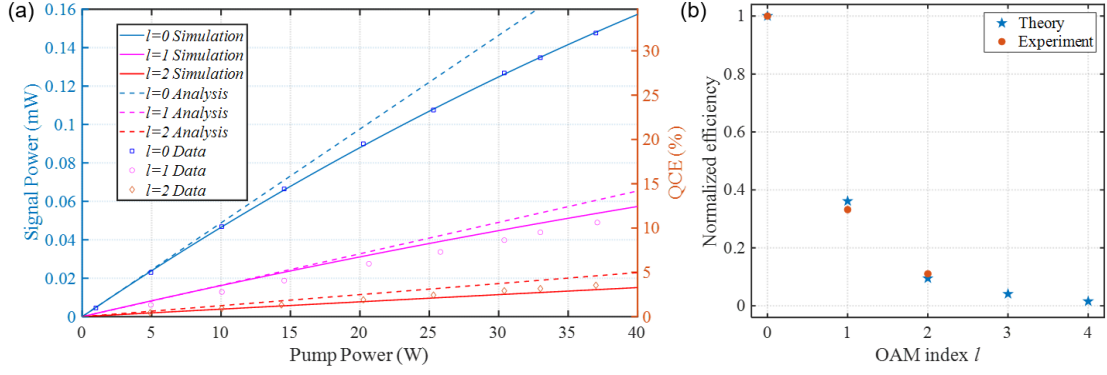

Fig. 3. Experimental results and theoretical predictions of the up-conversion process. (a) The relationships between the pump power and the SFG output powers for $l = 0, 1,$ and 2. The dashed lines are the analytical projections based on Eq. (3), while the solid lines present the results of numerical simulation. (b) Experimental results and predictions from numerical simulations of up-conversion efficiency for different OAM indexes.

During the two-step frequency transformation, the OAM should always be conserved as described in our previous study. Take the SFG process as an instance, assuming that the two input beams carried TC of $l_1$ and $l_2$ respectively, then the generated SFG light would have OAM of $(l_1 + l_2)\hbar$ [51]. In the DFG process, one of the input light carried the OAM with $l$, so the resulting mid-infrared and up-converted light should both carry OAM with the same TC. Based on the above theoretical analysis, the experiment result can be well explained now. The intensity distributions of the signal beam with $l = 1$ and 2, shown in Fig. 4 (a) and (c), were recorded by blocking one arm of the interferometer. And the output images of the interferometer in normal operation are shown in Fig. 4 (b) and (d). Fig. 4 (e-h) give the corresponding simulation results for Fig. 4 (a-d), exhibiting the same characteristics as the experimental results. The petal-like interference pattern shows the mode indices of the generated beam, just as we discussed in the preceding presentation. The numbers of petals in our experimental results indicated that the TCs of the up-converted light were 1 and 2 respectively, equaling to the TC of the original signal beam, which was in agreement with the theoretical prediction and numerical simulations.

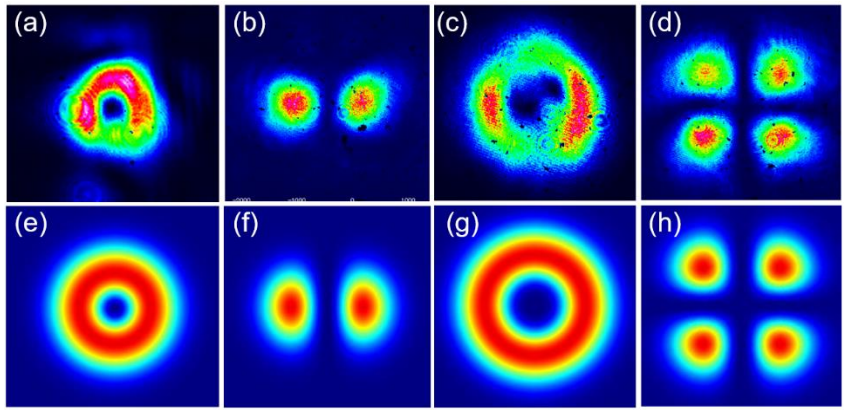

Fig. 4. The experimental results of the up-converted images. (a) and (c) The intensity distributions of up-converted light for $l = 1$ and 2. (b) and (d) The interference patterns for $l = 1$ and 2. (e)-(h) Images of the corresponding simulation results for (a)-(d).

The dependence of output power on the temperature of crystal for the DFG and SFG processes is shown in Fig. 5. The phase-matching temperatures were 55 °C and 55.2 °C, while the temperature bandwidths were 7.2°C and 6.9 °C, respectively. The experimental results of the power of the generated 3100-nm and 792-nm waves with different temperatures are marked in the figure, and the measured data can be fitted by solving the coupled wave equations [46]. The insert in the upper right corner of Fig. 4 shows how the phase mismatch affects the efficiency of SFG by numerical simulation, displaying a half peak width of 140 m$^{-1}$, while the same parameter given by the experimental conditions is 137 m$^{-1}$.

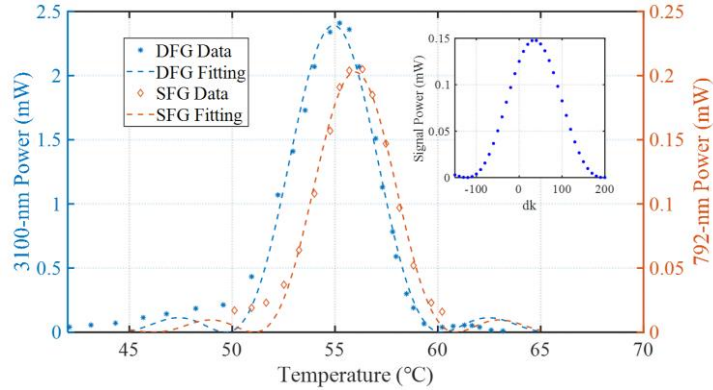

Fig. 5. The output power of DFG and SFG depending on temperature.

## 5. Conclusion

Based on our present experimental conditions, the intensity profile of the MIR beam could not be obtained directly. The length of the crystals and the internal nonuniformity affected the quality of generated beams to a certain extent, such as the generation of distortions and vortex splitting caused by aberration. Besides, the relatively small aperture of the crystal puts a limit on the choice of the focusing parameter, especially for the beam at a long wavelength. For up-conversion of image or light carrying OAM with higher-order TC, crystal with larger intersecting surface would perform better.

In summary, we have studied the frequency bridge between visible and MIR bands for vortex light based on QPM crystals. We generated the MIR beams through a DFG process and then demonstrated OAM frequency up-conversion experimentally for different OAM modes. The maximum quantum conversion efficiencies that were achieved for OAM modes with TCs of 0, 1, and 2 were 34.0%, 10.4%, and 3.5%, respectively. The experimental data were compared with the results of analytical expression and numerical simulation, proving the feasibility of theoretical prediction. We also verified the conservation of OAM in the cascaded processes and studied the dependence of the output power on the temperature of the crystals. The present work provides a reliable solution for up-conversion detection of light carrying OAM in the MIR band, using a bulk crystal that preserves phase information well. By adjusting the crystal parameters and reducing the noise, this setup could be quite hopefully extended to general image up-conversion detection and works at the single-photon level. This progress will be beneficial and encouraging for numerous applications that use MIR light as an information carrier and a means of detection, for example in the fields of biological detection, astronomical observation [56], environmental monitoring, and remote sensing.

## Acknowledgements

Project supported by the National Natural Science Foundation of China (NSFC) (92065101, 11934013); Anhui Initiative In Quantum Information Technologies (AHY020200).